\documentclass[sigconf]{acmart}
\renewcommand\footnotetextcopyrightpermission[1]{} 
\usepackage{amsmath,amssymb,amsfonts,bm,amsthm}
\usepackage[linesnumbered,ruled]{algorithm2e}
\usepackage{graphicx}
\usepackage{epsfig}
\usepackage[tight,normal]{subfigure}
\usepackage{multirow,makecell,url,footmisc}
\usepackage{enumerate}
\usepackage{textcomp}
\usepackage{booktabs} 
\usepackage{subfigure}
\usepackage{amssymb}
\usepackage{amsmath}
\usepackage{booktabs}
\usepackage{bm}
\usepackage{mathtools}
\usepackage{threeparttable}
\usepackage[american]{babel}
\usepackage{color}
\usepackage{threeparttable}
\usepackage{appendix}
\usepackage{comment}
\usepackage{multirow}
\usepackage{blindtext}
\usepackage{comment}
\usepackage{balance}

\usepackage{comment}

\def\BibTeX{{\rm B\kern-.05em{\sc i\kern-.025em b}\kern-.08emT\kern-.1667em\lower.7ex\hbox{E}\kern-.125emX}}

\begin{document}



\title{Search-based User Interest Modeling with Lifelong Sequential Behavior Data for Click-Through Rate Prediction}

\author{Pi Qi, Xiaoqiang Zhu, Guorui Zhou, Yujing Zhang, Zhe Wang, Lejian Ren, Ying Fan, and Kun Gai}
\authornote{} 
\affiliation{%
  \institution{Alibaba Group}
   \city{Beijing}
   \country{China}
}
\email{{piqi.pq, xiaoqiang.zxq, guorui.xgr, yujing.zyj, zhewang,wz, fanying.fy, lejian.rlj}@alibaba-inc.com, jingshi.gk@taobao.com
}

\renewcommand{\shortauthors}{}

\begin{abstract}
Rich user behavior data has been proven to be of great value for click-through rate prediction tasks, especially in industrial applications such as recommender systems and online advertising. Both industry and academy have paid much attention to this topic and propose different approaches to modeling with long sequential user behavior data. Among them, memory network based model MIMN\cite{pi2019deep} proposed by Alibaba, achieves SOTA with the co-design of both learning algorithm and serving system. MIMN is the first industrial solution that can model sequential user behavior data with length scaling up to 1000. However, MIMN fails to precisely capture user interests given a specific candidate item when the length of user behavior sequence increases further, say, by 10 times or more.  This challenge exists widely in previously proposed approaches. 


In this paper, we tackle this problem by designing a new modeling paradigm, which we name as  \textbf{S}earch-based \textbf{I}nterest \textbf{M}odel (SIM). SIM extracts user interests with two cascaded search units: (i) General Search Unit (GSU) acts as a general search from the raw and arbitrary long sequential behavior data, with query information from candidate item, and gets a \textbf{S}ub user \textbf{B}ehavior \textbf{S}equence (SBS) which is relevant to candidate item; (ii) Exact Search Unit (ESU) models the precise relationship between candidate item and SBS. This cascaded search paradigm enables SIM with a better ability to model lifelong sequential behavior data in both scalability and accuracy.  Apart from the learning algorithm, we also introduce our hands-on experience on how to implement SIM in large scale industrial systems. Since 2019, SIM has been deployed in the display advertising system in Alibaba, bringing 7.1\% CTR and 4.4\% RPM lift, which is significant to the business. Serving the main traffic in our real system now, SIM models sequential user behavior data with maximum length reaching up to 54000, pushing SOTA to 54x.

\end{abstract}




\maketitle
\section{Introduction}

Click-Through Rate (CTR) prediction modeling plays a critical role in industrial applications such as recommender systems and online advertising. Due to the rapid growth of user historical behavior data, user interest modeling, which focuses on learning the intent representation of user interest, has been widely introduced in the CTR prediction model \cite{zhou2018deep,zhou2019dien,youtube:recommend,pi2019deep}. However, most of the proposed approaches can only model sequential user behavior data with length scaling up to hundreds, limited by the burden of computation and storage in real online systems \cite{zhou2018deep,zhou2019dien}. Rich user behavior data is proven to be of great value \cite{pi2019deep}. For example, $23\%$ of users in Taobao, one of the world's leading e-commerce site, click with more than 1000 products in last 5 months\cite{pi2019deep,ren2019lifelong}. How to design a feasible solution to model the long sequential user behavior data has been an open and hot topic, attracting researchers from both industry and academy.  

A branch of research, which borrows ideas from the area of NLP, proposes to model long sequential user behavior data with memory network and makes some breakthroughs. MIMN\cite{pi2019deep} proposed by Alibaba, is one typical work, which achieves SOTA with the co-design of both learning algorithm and serving system. MIMN is the first industrial solution which can model sequential user behavior data with length scaling up to 1000. Specifically, MIMN incrementally embeds diverse interest of one user into a fixed size memory matrix which will be updated by each new behavior. In that way, the computation of user modeling is decoupled from CTR prediction. Thus for online serving, latency will not be a problem and the storage cost depends on the size of the memory matrix which is much less than the raw behavior sequence. A similar idea can be found in long-term interest modeling\cite{ren2019lifelong}. However, it is still challenging for memory network based approaches to model arbitrary long sequential data. Practically, we find that MIMN fails to precisely capture user interest given a specific candidate item when the length of user behavior sequence increases further, say, up to 10000 or more. This is because encoding all user historical behaviors into a fixed size memory matrix causes massive noise to be contained in the memory units.      
 
On the other hand, as pointed out in the previous work of DIN\cite{zhou2018deep}, the interest of one user is diverse and varies when facing different candidate items. The key idea of DIN is searching the effective information from user behaviors to model special interest of user, facing different candidate items. In this way, we can tackle the challenge of encoding all user interest into fixed-size parameters. DIN does bring a big improvement for CTR modeling with user behavior data. But the searching formula of DIN costs an unacceptable computation and storage facing the long sequential user behavior data as we mentioned above. So, can we apply a similar search trick and design a more efficient way to extract knowledge from the long sequential user behavior data? 

In this paper, we tackle the challenge by designing a new modeling paradigm, which we name as  \textbf{S}earch-based \textbf{I}nterest \textbf{M}odel (SIM). SIM adopts the idea of DIN \cite{zhou2018deep} and captures only relevant user interest with respect to specific candidate items. In SIM, user interest is extracted with two cascaded search units:(i) General Search Unit (GSU) acts as a general search from the raw and arbitrary long sequential behavior data, with query information from candidate item, and gets a \textbf{S}ub user \textbf{B}ehavior \textbf{S}equence (SBS) which is relevant to candidate item. In order to meet the strict limitation of latency and computation resources, general but effective methods are utilized in the GSU. To our experience, the length of SBS can be cut down to hundreds and most of the noise information in raw long sequential behavior data could be filtered. (ii) Exact Search Unit (ESU) models the precise relationship between the candidate item and SBS. Here we can easily apply similar methods proposed by DIN\cite{zhou2018deep} or DIEN\cite{zhou2019dien}. 


The main contributions of this work are summarized as follows:
\begin{itemize}
\item We propose a new paradigm SIM for modeling long sequential user behavior data. The design of a cascaded two-stage search mechanism enables SIM with a better ability to model lifelong sequential behavior data in both scalability and accuracy.  
\item We introduce our hands-on experience of implementing SIM in large scale industrial systems. Since 2019, SIM has been deployed in the display advertising system in Alibaba, bringing $7.1\%$ CTR and $4.4\%$ RPM lift. Now, SIM is serving the main traffic.
\item We push the maximum length for modeling with long sequential user behavior data up to 54000, 54x larger than MIMN, the published SOTA industry solution for this task.    
\end{itemize}

\section{Related work}
\textbf{User Interest Model.} Deep learning based methods have achieved great success in CTR prediction task\cite{cheng2016wide,deep_intent,deep_crossing}. In early age, most pioneer works\cite{guo2017deepfm,cheng2016wide,lian2018xdeepfm,qu2016product-based,wang2017deep} use a deep neural network to capture interactions between features from different fields so that engineers could get rid of boring feature engineering works. Recently, a series of works, which we called User Interest Model, focus on learning the representation of latent user interest from historical behaviors, using different neural network architecture such as CNN\cite{tang2018personalized,yuan2019simple}, RNN\cite{hidasi2015session,zhou2019dien}, Transformer\cite{sun2019bert4rec,feng2019deep} and Capsule\cite{li2019multi}, etc. DIN\cite{zhou2018deep} emphasizes that user interest are diverse and an attention
mechanism is introduced in DIN to capture users' diverse interest on the different target items. DIEN\cite{zhou2019dien} points out that the temporal relationship between historical behaviors matters for modeling users' drifting interest. An interest extraction layer based on GRU with auxiliary loss is designed in DIEN. MIND\cite{li2019multi} argues that using a single vector to represent one user is
insufficient to capture the varying nature of the user's interest. Capsule network and dynamic routing method are introduced in MIND to learn the representation of user interest as multiple vectors. Moreover, inspired by the success of the self-attention architecture in the tasks of sequence to sequence learning, Transformer is introduced in \cite{feng2019deep} to model user cross-session and in-session interest. 

\textbf{Long-term User Interest.} \cite{pi2019deep} shows that considering long-term historical behavior sequences in the user interest model can significantly improve CTR model performance. Although longer user behavior sequences bring in more useful information for user interest modeling, it extremely burdens the latency and storage of an online serving system and contains massive noise for point-wise CTR prediction at the same time. A series of works focus on tackling the challenge in long-term user interest modeling, which usually learns user interest representation based on historical behavior sequences with extremely large length even lifelong. \cite{ren2019lifelong} proposes a Hierarchical Periodic Memory Network for lifelong sequential modeling
with personalized memorization of sequential patterns for each
user. 
\cite{yu2019adaptive} choose an attention-based framework to combine users' long-term and short-term preferences. And they adopt the attentive Asymmetric-SVD paradigm to model long-term interest.
A memory-based architecture named MIMN is proposed in \cite{pi2019deep}, which embedded user long-term interest into fixed-sized memory network to solve the problem of large storage of user behavior data. And a UIC module is designed to record the new user behaviors incrementally to deal with the latency limitation.
But MIMN abandons the information from the target item in the memory network which has been proved to be important for user interest modeling.

\section{Search-based Interest Model}

\begin{figure*}[t]
    \centering
    \includegraphics[height=3in, width=6.0in]{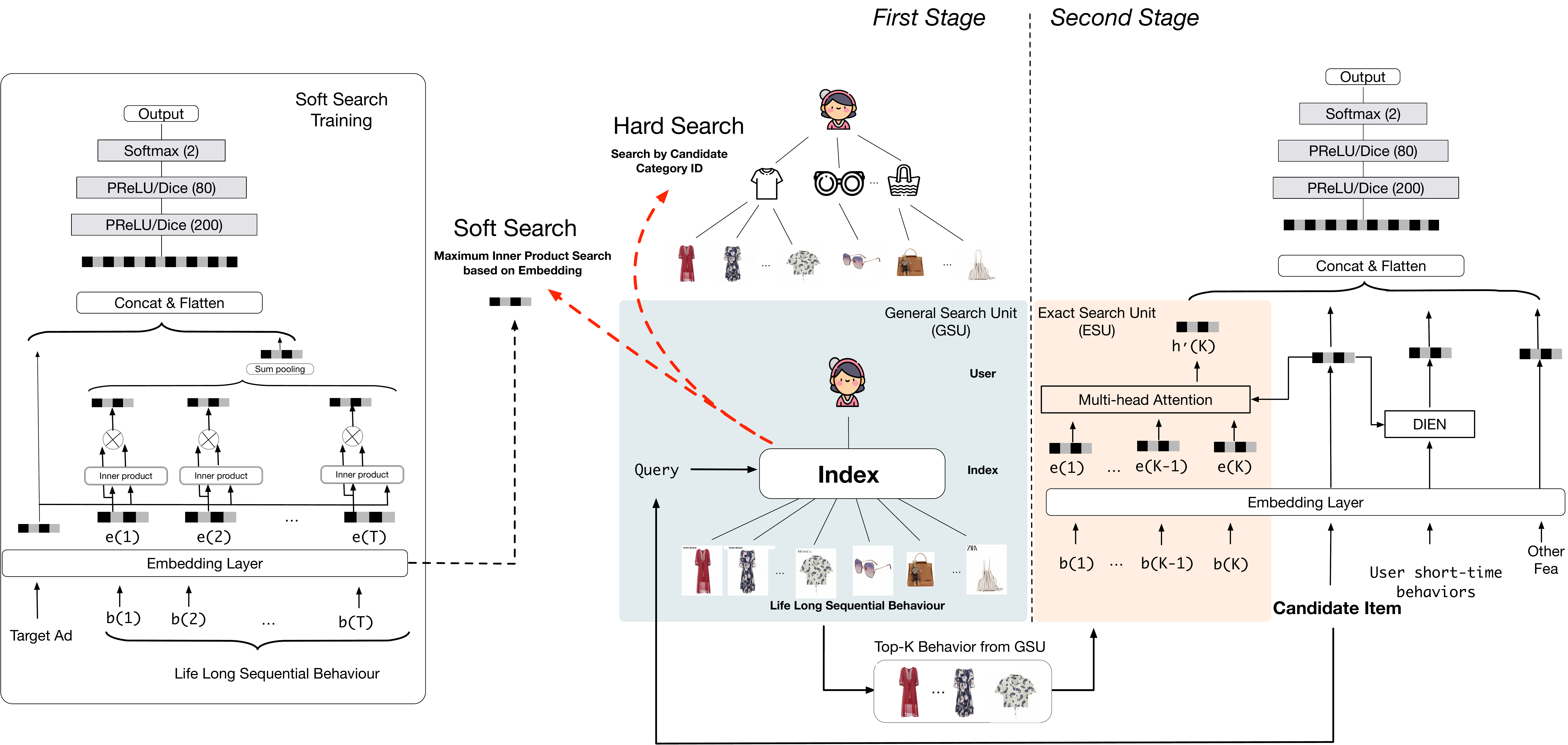} 
    \caption{Search-based Interest Model (SIM). SIM follows a two-stage search strategy and is composed of two units: (i) the General Search Unit seeks the most related K behaviors from over ten thousand user behaviors, (ii) the Exact Search Unit utilizes multi-head attention to capture the diverse user interest. And then it follows the traditional Embedding\&MLP paradigm which takes the output of precise long-time user interest and other features as inputs. 
   In this paper, we introduce hard-search and soft-search for the GSU. Hard-search means select the behavior data belongs to the same category of the candidate item. Soft-search means indexing each user behavior data based on the embedding vectors and using maximum inner product search to seek Top-K behavior. For soft-search, GSU and ESU share the same embedding params which are trained simultaneously during the learning process, and the Top-K behavior sequence is generated based on the newest params.} 
    \label{fig:SIM}
   \end{figure*}

It has been proven to be effective by modeling user behavior data for CTR prediction modeling. 
Typically, attention-based CTR models, such as DIN\cite{zhou2018deep} and DIEN\cite{zhou2019dien}, design complex model structure and involve attention mechanism to capture user diverse interest by searching effective knowledge from user behavior sequence, with inputs from different candidate items. But in a real-world system, these models can only handle short-term behavior sequence data, of which the length is usually less than 150. On the other hand, the long-term user behavior data is valuable, and modeling the long-term interest of users may bring more diverse recommendation results for users. It seems we are on the horns of a dilemma: we cannot handle the valuable life-long user behavior data with the effective but complex methods in a real-world system. 

To tackle this challenge, we  propose a new modeling paradigm, which is named as  \textbf{S}earch-based \textbf{I}nterest \textbf{M}odel (SIM). SIM follows a two-stage search strategy and can handle long user behavior sequences in an efficient way. In this section, we will introduce the overall workflow of SIM first and then introduce the two proposed search units in detail.


\subsection{Overall Workflow}
The overall workflow of SIM is shown in Figure \ref{fig:SIM}. SIM follows a cascaded two-stage search strategy with two corresponding units: the General Search Unit (GSU) and the Exact Search Unit (ESU).

\textbf{In the first stage}, we utilize General Search Unit (GSU) to seek top-K relevant sub behavior sequences from original long-term behavior sequences with sub-linear time complexity. Here K is generally much shorter than the original length of behavior sequences. An efficient search method can be conducted if the relevant behaviors can be searched under the limitation of time and computation resources. In section \ref{sec:GSU}  
we provide two straightforward implementations of GSU: soft-search and hard-search. GSU takes a general but effective strategy to cut off the length of raw sequential behaviors to meet the strict limitation of time and computation resources. Meanwhile, the massive noise that exists in the long-term user behavior sequence, which may undermine user interest modeling, can be filtered by the search strategy in the first stage.  

\textbf{In the second stage}, the Exact Search Unit (ESU), which takes the filtered sub-sequential user behaviors as input, is introduced to further capture the precise user interest.  
Here a sophisticated model with complex architecture can be applied, such as DIN\cite{zhou2018deep} and DIEN\cite{zhou2019dien}, as the length of long-term behaviors has been reduced to hundreds.  

Note that although we introduce the two stages separately, actually they are trained together.

\subsection{General Search Unit}
\label{sec:GSU}
Given a candidate item (the target item to be scored by CTR model), only a part of user behaviors are valuable. This part of user behaviors are closely related to final user decision. Picking out these relevant user behaviors is helpful in user interest modeling. However, using the whole user behavior sequence to directly model the user interest will bring enormous resource usage and response latency, which is usually unacceptable in practical applications. 
To this end, we propose a general search unit to cut down the input number of user behaviors in user interest modeling. Here we introduce two kinds of general search unit: hard-search and soft-search. 

Given the list of user behaviors $\mathbf{B}=[\mathbf{b}_{1};\mathbf{b}_{2};\cdots;\mathbf{b}_{T}]$, where $\mathbf{b}_{i}$ is the $i$-th user behavior and $T$ is the length of user behaviors. The general search unit calculate relevant score $r_i$ for each behavior $\mathbf{b}_{i}$ w.r.t. the candidate item and then select the Top-K relevant behaviors with score $r_i$ as sub behaviour sequence $\mathbf{B}^*$. The difference between hard-search and soft-search is the formulation of relevant score $r_i$: 
\begin{equation}
    r_i = \left\{
    \begin{aligned}
        &Sign(C_i = C_a) & & hard-search\\
        &(W_{b}\mathbf{e}_{i}) \odot (W_{a}\mathbf{e}_a)^T & & soft-search 
    \end{aligned}
    \right.
\end{equation}

\textbf{Hard-search}. The hard-search model is non-parametric. Only behavior belongs to the same category as the candidate item will be selected and aggregated as a sub behavior sequence to be sent to the exact search unit. Here $C_a$ and $C_i$ denote the categories of target item and the $i$-th behavior $b_i$ that belong to correspondingly. Hard-search is straightforward but later in section 4 we will show it is quite suitable for online serving.    
 
\textbf{Soft-search}.  In the soft-search model,  $\mathbf{b}_{t}$ is first encoded as one-hot vector and then embedded into low-dimensional vectors $\mathbf{E}=[\mathbf{e}_{1};\mathbf{e}_{2};\cdots;\mathbf{e}_{T}]$, as shown in Figure \ref{fig:SIM}. $W_{b}$ and $W_{a}$ are the parameters of weight. $\mathbf{e}_a$ and $\mathbf{e}_i$ denote the embedding vectors of target item and $i$-th behavior $b_i$, respectively. To further speed up the top-K search over ten thousands length of user behaviors, sub-linear time maximum inner product search method ALSH\cite{shrivastava2014asymmetric} is conducted based on the embedding vectors $\mathbf{E}$ to search the related top-K behaviors with target item. With the well-trained embedding and Maximum Inner Product Search (MIPS) method, over ten thousands user behaviors could be reduced to hundreds. 

It should be noticed that distributions of long-term and short-term data are different. Thus, directly using the parameters learned from short-term user interest modeling in soft-search model may mislead the long-term user interest modeling. In this paper, the parameters of soft-search model is trained under an auxiliary CTR prediction task based on long-term behavior data, illustrated as soft search training in the left of Figure \ref{fig:SIM}. The behaviors representation $\mathbf{U}_{r}$ is obtained by multiplying the $r_i$ and $\mathbf{e}_i$: 
\begin{align}
&\mathbf{U}_r= \sum_{i=1}^T r_i \mathbf{e}_{i}.
\end{align}
The behaviors representation $\mathbf{U}_{r}$ and the target vector $\mathbf{e}_{a}$ are then concatenated as the input of following MLP (Multi-Layer Perception). Note that if the user behavior grows to a certain extent, it is impossible to directly fed the whole user behaviors into the model. In that situation, one can randomly sample sets of sub-sequence from the long sequential user behaviors, which still follows the same distribution of the original one.

\subsection{Exact Search Unit}
\label{sec:ESU}
In the first search stage, top-$K$ related sub user behavior sequence $\mathbf{B}^*$ w.r.t. the target item is selected from long-term user behaviors. To further model the user interest from the relevant behaviors, we introduce the Exact Search Unit  which is an attention-based model taking $\mathbf{B}^*$ as input.

Considering that these selected user behaviors across a long time so that the contribution of user behaviors are different, we involve the sequence temporal property for each behavior. 
Specifically,the time intervals $\mathbf{D} = [\Delta_1; \Delta_2; ...; \Delta_K]$ between target item and selected $K$ user behaviors are used to provide temporal distance information. The $\mathbf{B}^*$ and $\mathbf{D}$ are encoded as embedding $\mathbf{E}^* = [\mathbf{e}_{1}^*;\mathbf{e}_{2}^*; ...; \mathbf{e}_{K}^*]$ and $\mathbf{E}_\mathbf{t}=[\mathbf{e}^t_{1};\mathbf{e}^t_{2}; ...; \mathbf{e}^t_{K}]$, respectively. $\mathbf{e}_j^*$ and $\mathbf{e}_j^t$ are concatenated as the the final representation of the user behavior which is denoted as $\mathbf{z}_{j}=concat(\mathbf{e}_{j}^*,\mathbf{e}^t_{j})$. We take advantage of multi-head attention to capture the diverse user interest:
\begin{align}
&\mathbf{att}^{i}_{score}= Softmax(W_{bi}\mathbf{z}_{b} \odot \ W_{ai}\mathbf{e}_a) \\
&\mathbf{head}_{i} = \mathbf{att}^{i}_{score} \mathbf{z}_{b},
\end{align}
where $\mathbf{att}^{i}_{score}$ is the $i$-th attention score and $\mathbf{head}_{i}$ is the $i$-th head in multi-head attention. $W_{bi}$ and $W_{ai}$ are the $i$-th parameter of weight. The final user longtime diverse interest is represent as $U_{lt}=concat(head_{1}; ...; head_{q})$. It is then fed into the MLP for CTR prediction.

Finally, the general search unit and exact search unit are trained simultaneously under Cross-entropy loss function. 
\begin{equation}
    Loss = \alpha Loss_{GSU} + \beta Loss_{ESU},
\end{equation}
where $\alpha$ and $\beta$ are hyper parameters to control the loss weights. In our experiments, if GSU use soft-search model, both $\alpha$ and $\beta$ are set as $1$. GSU with hard-search model is nonparametric and the $\alpha$ is set as $0$.

\begin{figure}[t]
 \centering
 \includegraphics[height=2in,width=3.2in]{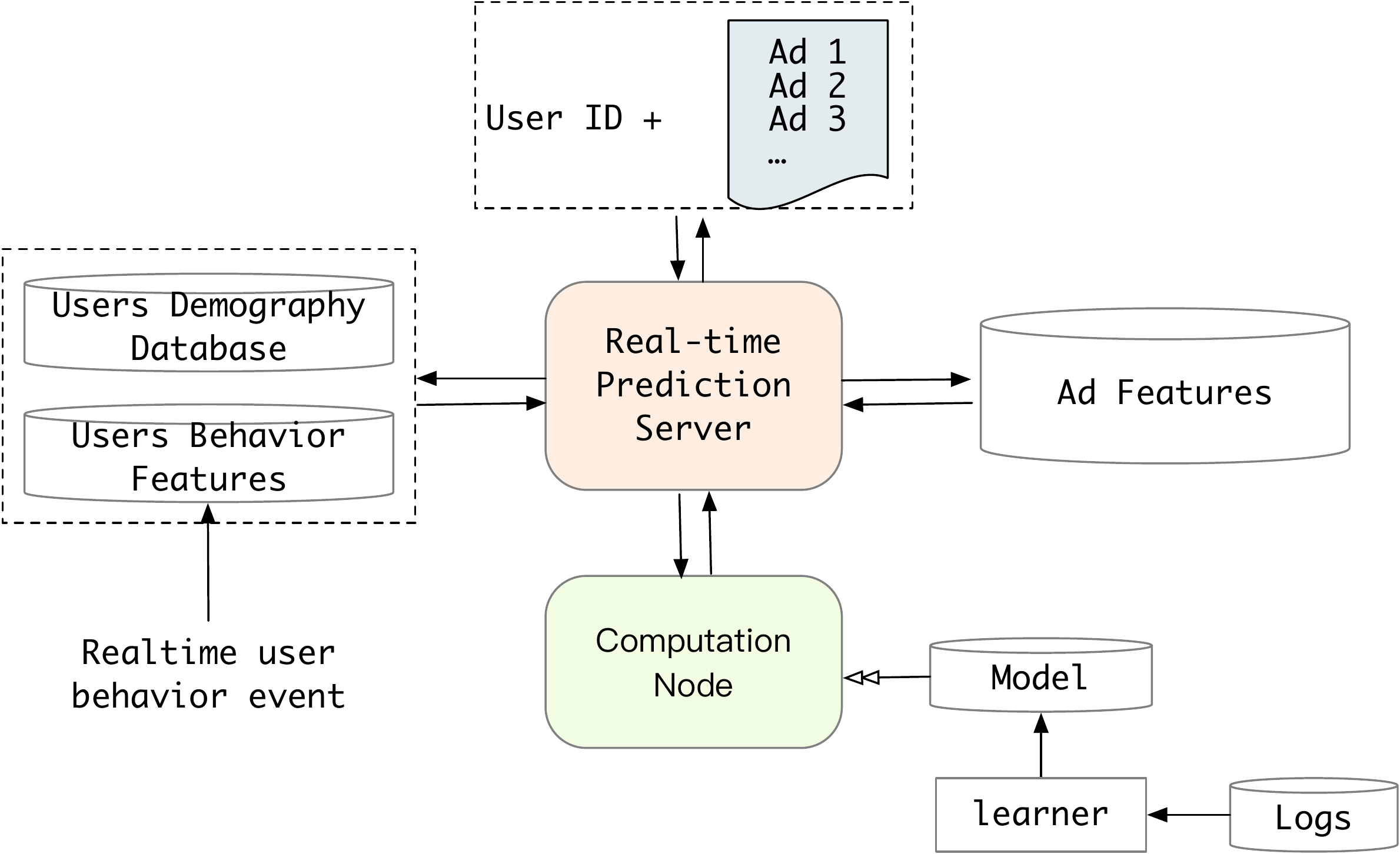} 
 \caption{Real-Time Prediction (RTP) system for CTR task in our industrial display advertising system. It consists of two key components: computation node and prediction server. 
 It would bring great pressure of storage and latency for RTP online system with long sequential user behavior data.}   
 \label{fig:system}
\end{figure}

\section{Implementation for Online Serving} 
\label{sec:bg}

In this section, we introduce our hands-on experience of implementing SIM in the display advertising system in Alibaba.  

\subsection{Challenges of Online Serving with Lifelong User Behavior Data}
\label{sec:challenge}
Industrial recommender or advertising systems need to process massive traffic requests in one second, which needs the CTR model to response in real-time. 
Typically, the serving latency should be less than 30 milliseconds. Figure \ref{fig:system} briefly illustrates \textbf{R}eal \textbf{T}ime \textbf{P}rediction (\textbf{RTP}) system for CTR task in our online display advertising system. 

Taking lifelong user behavior into consideration, it's even harder to make a long-term user interest model serving in real-time industrial system. The \textbf{storage} and \textbf{latency} constraints could be the bottleneck of the long-term user interest model\cite{pi2019deep}. Traffic would increase linearly as the length of user behavior sequence grows. Moreover, our system serves more than 1 million users per second at traffic peak. Hence, it is a great challenge to deploy a long-term model to the online system.


\subsection{Search-based Interest Model for Online Serving System}
\label{sec:sim}
In section \ref{sec:GSU}, we propose two kinds of general search unit, soft-search model and hard-search model. For both soft and hard search model, we conduct extensive offline experiments on industrial data, which is collected from the online display advertising system in Alibaba. We observe that the Top-K behavior generated from soft-search model is extremely similar to the result of hard-search model. In other words, most of the top-K behavior from soft-search generally belong to the category of the target item. It is a characteristic of data in our scenario. In e-commerce website, items belong to the same category are similar in most cases. 
Considering this, although soft-search model performs slightly better than hard-search model in offline experiments, refer to table 4 for details, after balancing the performance gain and resource consumption, we choose the hard-search model to deploy SIM in our advertising system.

\begin{figure}[t]
 \centering
 \includegraphics[height=2in, width=3.5in]{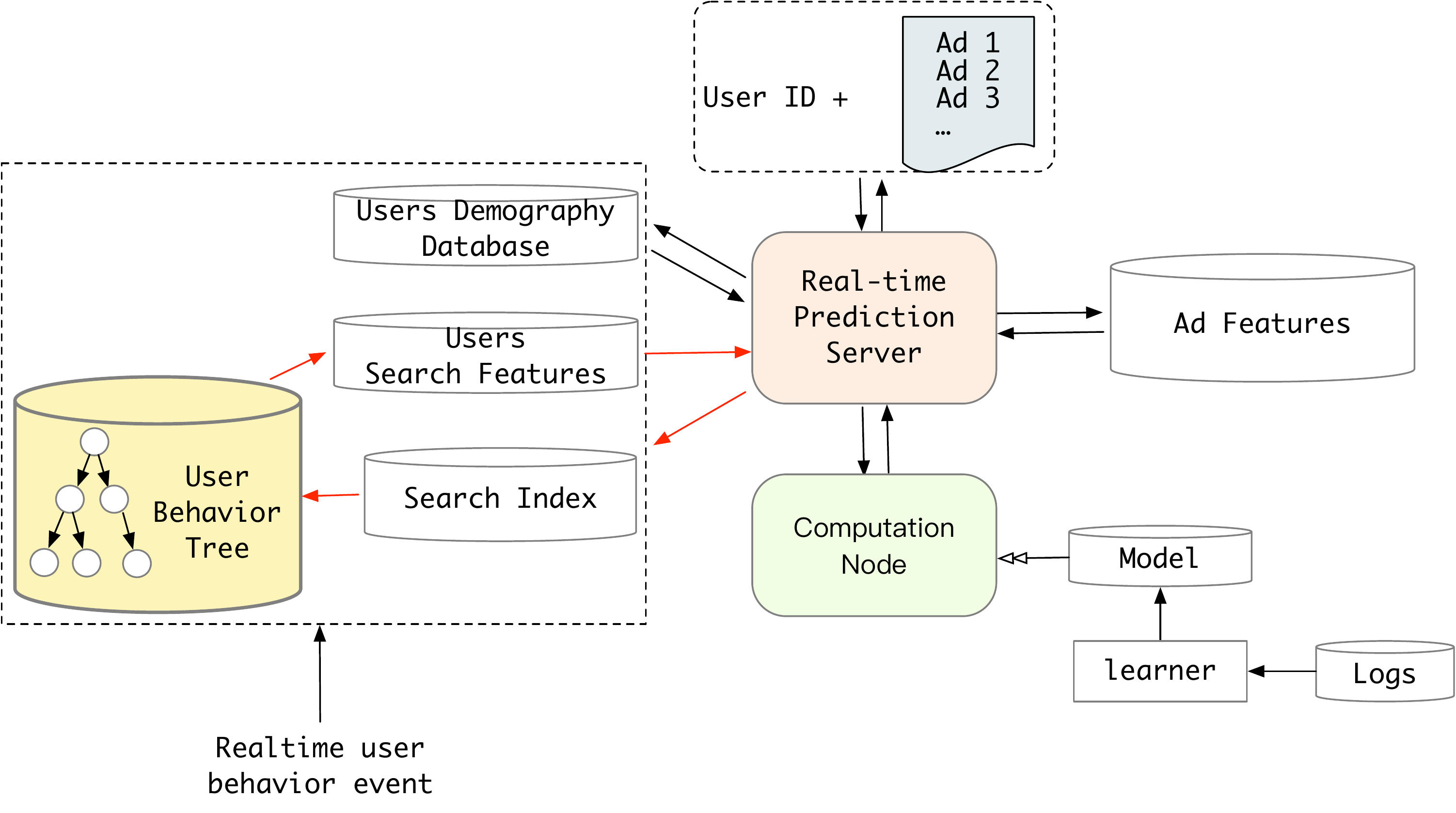} 
 \caption{CTR prediction system with proposed SIM model. 
 The new system joins a hard-search module to seek the effective behaviors with target item from long sequential behavior data. The index of user behavior tree is built early in the offline manner, saving most of latency cost for online serving.}    
 \label{fig:sim_system}
\end{figure}

For hard-search model, the index which contains all the long sequential behavior data is a key component. We observe that  behaviors can be achieved naturally by the category they belong to. Hence, we build an two-level structured index for each user, which we name as user behavior tree (UBT), as illustrated in Figure \ref{fig:sim_system}. Briefly speaking, UBT follows the Key-Key-Value data structure: the first key is user id, the second keys are category ids and the last values are the specific behavior items that belong to each category. UBT is implemented as an distributed system, with size reaching up to 22 TB, and is flexible enough to provide high throughput query. Then, we take the category of target item as our hard-search query.  After the general search unit, the length of user behaviors could be reduced from over ten thousands to hundred. Thus, the storage pressure of lifelong behaviors in online system could be released. Figure \ref{fig:sim_system} shows the new CTR prediction system with search-based interest model. 

Note that the index of user behavior tree for the general search unit can be pre-built offline. In that way, the response time for general search unit in online system could be really short and can be omitted comparing to the calculation of GSU. Besides, other user features can be computed in parallel. 

 
\section{Experiments}
In this section, we present our experiments in detail, including
datasets, experimental setup, model comparison, and some corresponding analyses. The proposed search model is compared with several state-of-the-art work on two public datasets and one industrial dataset as shown in Table \ref{table:Statistics}.
Since SIM has been deployed in our online advertising system, we also conduct careful online A/B testing, with a comparison of several famous industry models. 

\subsection{Datasets}
Model comparisons are conducted on two public datasets as well as an industrial dataset collected from the online display advertising system of Alibaba. Table \ref{table:Statistics} shows the statistics of all datasets.  

\begin{table}[]
\caption{Statistics of datasets used in this paper.}
\small
\centering
\begin{threeparttable}
\begin{tabular}{lcccc}
\toprule
    Dataset       & Users & Items\tnote{a} & Categories & Instances \\ \midrule
Amazon (Book). & 75053 & 358367 & 1583 & 150016 \\ 
Taobao. & 7956431 & 34196612 & 5597 & 7956431 \\
Industrial.  & 0.29 billion & 0.6 billion & 100,000 & 12.2 billion \\ 
\bottomrule
\end{tabular}
            \begin{tablenotes}
            \item[a] For industrial dataset, items refer to be the advertisements.
            \end{tablenotes}
 \end{threeparttable}
\label{table:Statistics}
\end{table} 

\textbf{Amazon Dataset\footnote{http://jmcauley.ucsd.edu/data/amazon/}} is composed of product reviews and metadata from Amazon. We use the \textsl{Books} subset of the Amazon dataset which contains 75053 users, 358367 items, and 1583 categories. For this dataset, we regard reviews as one kind of interaction behaviors and sort the reviews from one user by time. The maximum behavior sequence length of the amazon book dataset is 100. We split the recent 10 user behaviors as short-term user sequential features and recent 90 user behaviors as long-term user sequential features. These pre-processing method has been widely used in related works.

\textbf{Taobao Dataset} is a collection of user behaviors from Taobao's recommender system. The dataset contains several types of user behaviors including click, purchase, etc. It contains user behavior sequences of about eight million users. We take the click behaviors for each user and sort them according to time in an attempt to construct the behavior sequence.  The maximum behavior sequence length of the Taobao dataset is 500. We split the recent 100 user behaviors as short-term user sequential features and the recent 400 user behaviors as long-term user sequential features.  The dataset will be published soon.

\textbf{Industrial Dataset} is collected from the online display advertising system of Alibaba. Samples are constructed from impression logs, with ``click'' or ``not'' as the label. The training set is composed of samples from the past 49 days and test set from the following day, a classic setting for industrial modeling. In this dataset, user behavior feature in each day's sample contains historical behavior sequences from the preceding $180$ days as long-term behavior features and that from the preceding $14$ days as short-term behavior features. Over 30\% of samples contain sequential behavior data with a length of more than 10000. Moreover, the maximum length of behavior sequence reaches up to 54000, which is 54x larger than that in MIMN \cite{pi2019deep}. 

\subsection{Competitors and experiment setup}
We compare SIM with mainstream CTR prediction models as follows. 
\begin{itemize}
        \item {\textbf{DIN}} \cite{zhou2018deep} is an early work for user behavior modeling which proposes to soft-search user behaviors w.r.t. candidates. Compared with other long-term user interest model, DIN only takes short-term user behaviors as input.      
        \item{ \textbf{Avg-Pooling Long DIN}} To compare model performance on long-term user interest, we applied average-pooling operation on long-term behavior and concatenate the long-term embedding with other feature embeddings.
        \item{ \textbf{MIMN}} \cite{pi2019deep} which has ingeniously designed model architecture to capture long-term user interest achieves state-of-art performance. 
        \item{ \textbf {SIM (hard)}} is the  proposed SIM model with hard-search in first stage without time embedding in ESU.
        \item{ \textbf{SIM (soft)}} is the proposed SIM model with soft-search in the first stage without time embedding in ESU.
        \item{ \textbf{SIM (hard/soft) with Timeinfo}} is SIM with hard/soft search in first stage with time embedding. 
\end{itemize}

\textbf{Experiment Settings.} We take the same experiment setup with related works\cite{pi2019deep} so that the experiment results can be compared fairly. For all models, we use Adam solver.
We apply exponential decay with the learning rate starting at 0.001. Layers of fully connected network (FCN)
are set by $200 \times 80 \times 2$. The number of embedding dimension is set to be 4. And we take widely used AUC as model performance measurement metrics.

\subsection{Results on Public Datasets}
\label{sec:result_public}
\begin{table}
    \small
    \centering
    \caption{Model performance  (AUC) on public datasets}\label{tab:public}
    \begin{threeparttable}
        \begin{tabular}{l c c}
            \addlinespace
            \toprule
            Model        & Taobao  (mean $\pm$ std) & Amazon  (mean $\pm$ std)\\
            \midrule        
            {DIN}~ & $ 0.9214 \pm 0.00017$  &  $ 0.7276 \pm 0.00051 $  \\
            {Avg-Pooling Long DIN}~    & $ 0.9281 \pm 0.00025$  &  $ 0.7280 \pm 0.00012 $  \\
            {MIMN}~    & $ 0.9278 \pm 0.00035$  &  $ 0.7396 \pm 0.00037 $  \\
            {SIM (soft)\tnote{a}}~    & $0.9416 \pm 0.00049$  &  $\bm{0.7510 \pm 0.00052}$  \\
            {SIM (soft) with Timeinfo}~    & $\bm{0.9501 \pm 0.00017}$  &  -\tnote{b}  \\
            \bottomrule
    \end{tabular}
    \begin{tablenotes}
    \item[a] SIM (soft) is SIM with soft search without time interval embeddings
    \item[b] We didn't conduct the experiment on Amazon Dataset, as there are no timestamp features in it
    \end{tablenotes}
    \end{threeparttable}
\end{table}
Table~\ref{tab:public} presents the results of all the compared models. Compared with  DIN, the other models that take in long-term user behavior features perform much better. It demonstrates that long-term user behavior is helpful for CTR prediction task. 
SIM achieves significant improvement compared with MIMN because MIMN encodes all unfiltered user historical behaviors into a fixed-length memory which makes it hard to capture diverse long-term interest. SIM uses a two-stage search strategy to search relevant behaviors from a huge massive historical sequential behaviors and models the diverse long-term interest vary on different target items. Experiment results show that SIM outperforms all the other long-term interest model which strongly proves that the proposed two-stage search strategy is useful for long-term user interest modeling. Moreover, involving time embedding could achieve further improvement.



\begin{table}
    \caption{Effectiveness evalutaion of two-stage search architecture on longterm user interest modeling}\label{tab:stage}
    \resizebox{\columnwidth}{!}{%
        \begin{tabular}{l c c}
            \addlinespace
            \toprule
            Operations ~ & Taobao  (mean $\pm$ std) & Amazon  (mean $\pm$ std)\\
            \midrule        
            {Avg-Pooling without Search}~ & $ 0.9281 \pm 0.00025$  &  $ 0.7280 \pm 0.00012 $  \\
            {Only First Stage (hard)}~    & $ 0.9330 \pm 0.00031$  &  $ 0.7365 \pm 0.00022 $  \\
            {Only First Stage (soft)}~    & $ 0.9357 \pm 0.00025$  &  $ 0.7342 \pm 0.00012 $  \\
            {SIM (hard)}~    & $\bm{0.9332 \pm 0.00008}$  &  $\bm{0.7413 \pm 0.00016}$  \\
            {SIM (soft)}~    & $\bm{0.9416 \pm 0.00049}$  &  $\bm{0.7510 \pm 0.00052}$  \\
            {SIM (soft) with Timeinfo}~    & $\bm{0.9501 \pm 0.00017}$  &  -  \\
            \bottomrule
    \end{tabular}
    }
\end{table}

\begin{table}
        \centering
        \caption{Model performance (AUC) on industrial dataset}
        \label{tab:deploy}
        \begin{threeparttable}
                \begin{tabular}{p{5cm} p{1.5cm}<{\centering}}
                        \addlinespace
                        \toprule
                     ~~~~~   Model ~~~~~  &AUC      ~~~~~ \\
                        \midrule
                    ~~~~~   {DIEN            } ~~~~~& $0.6452$  ~~~~~\\
                     ~~~~~   {MIMN            } ~~~~~& $0.6541$  ~~~~~\\
                     ~~~~~   {SIM (hard)            } ~~~~~& $0.6604$     ~~~~~ \\
                      ~~~~~   {SIM (soft)          } ~~~~~& $0.6625$     ~~~~~ \\
                     ~~~~~   {SIM (hard) with timeinfo\tnote{a}}     ~~~~~&$0.6624 $ ~~~~~ \\
                     \bottomrule
        \end{tabular}
        \begin{tablenotes}
        \item[a] The model has been deployed in our online serving system and is serving the main traffic now.
        \end{tablenotes}
        \end{threeparttable}
\end{table}

\subsection{Ablation Study}
\textbf{Effectiveness of the two-stage search.} As mentioned above, the proposed search interest model uses a two-stage search strategy. 
The first stage follows a general search strategy to filter out relevant historical behaviors with the target item.
The second stage conducts an attention-based exact search on behaviors from the first stage to accurately capture users' diverse long-term interest on target items.
 In this section, we will evaluate the effectiveness of the proposed two-stage search architecture by experiments with different operations applied to long-term historical behavior. As shown in Table~\ref{tab:stage}, {\it Avg-Pooling without Search} is simply using average pooling to integrate long-term behavior embedding without any filters, same as Avg-pooling Long DIN. {\it Only First Stage(hard)} is applying hard-search on long-term historical behaviors in the first stage and integrate filtered embedding by average pooling to a fixed size vector as the input of MLP. {\it Only First Stage (soft)} is almost the same as {\it Only First Stage (hard)} except applying parametric soft-search rather than hard-search at the first stage. In the third experiment, we offline calculate the inner product similarity score between a target item and long-term user behaviors based on pre-trained embedding vectors. The soft-search is conducted by selecting the top 50 relevant behaviors according to the similarity score. 
 And the last three experiments are the proposed search model with two-stage search architecture.

As shown in Table~\ref{tab:stage}, all the methods with filter strategy extremely improve model performance compared with simply average pooling the embedding. It indicates that there indeed exists massive noise in original long-term behavior sequences which may undermine long-term user interest learning. Compared with models with only one stage search. The proposed search model with a two-stage search strategy makes further progress by introducing an attention-based search on the second stage. It indicates that precisely modeling users' diverse long-term interest on target items is helpful on CTR prediction tasks. And after the first stage search, the filtered behavior sequences usually are much shorter than the original sequences. The attention operations wouldn't burden the online serving RTP system too much.

Involving time embedding achieves further improvement which demonstrates that the contribution of user behaviors in different period differs.

\subsection{Results on Industrial Dataset}
\label{sec:result_industrial}
We further conduct experiments on the dataset collected from the online display advertisement system of Alibaba.  Table~\ref{tab:deploy} shows the results. Compared with hard-search in the first stage, soft-search performs better. Meanwhile, we notice that there is just a slight gap between the two search strategies at the first stage. Applying the soft-search in the first stage cost more computation and storage resources. Because the nearest neighbors search method would be utilized in online serving, while hard-search only need searching from a two-level index table which would be built in offline. 
Hence hard-search is more efficient and system friendly. Moreover, for two different search strategies, we do statistics on over 1 million samples and 100 thousand users with long-term historical behaviors from the industrial dataset. The result shows that the user behaviors reserved by hard-search strategy could cover 75\% of that from the soft-search strategy. Finally, we choose the simpler hard-search strategy at the first stage of the trade-off between efficiency and performance. SIM improves MIMN with AUC gain of 0.008, which is significant for our business.

\textbf{Online A/B Testing.} Since 2019, we have deployed the proposed solution in the display advertising system in Alibaba.
From 2020-01-07 to 2020-02-07, we conduct a strict online A/B testing experiment to validate the proposed SIM model. Compared to MIMN (our last product model), SIM achieves great gain in Alibaba display advertising scene, which shows in table \ref{tab:online}. Now, SIM has been deployed online and serves the main scene traffic every day, which contributes significant business revenue growth.


\begin{figure}[t]
    \centering
    \includegraphics[height=1.9in, width=3.5in]{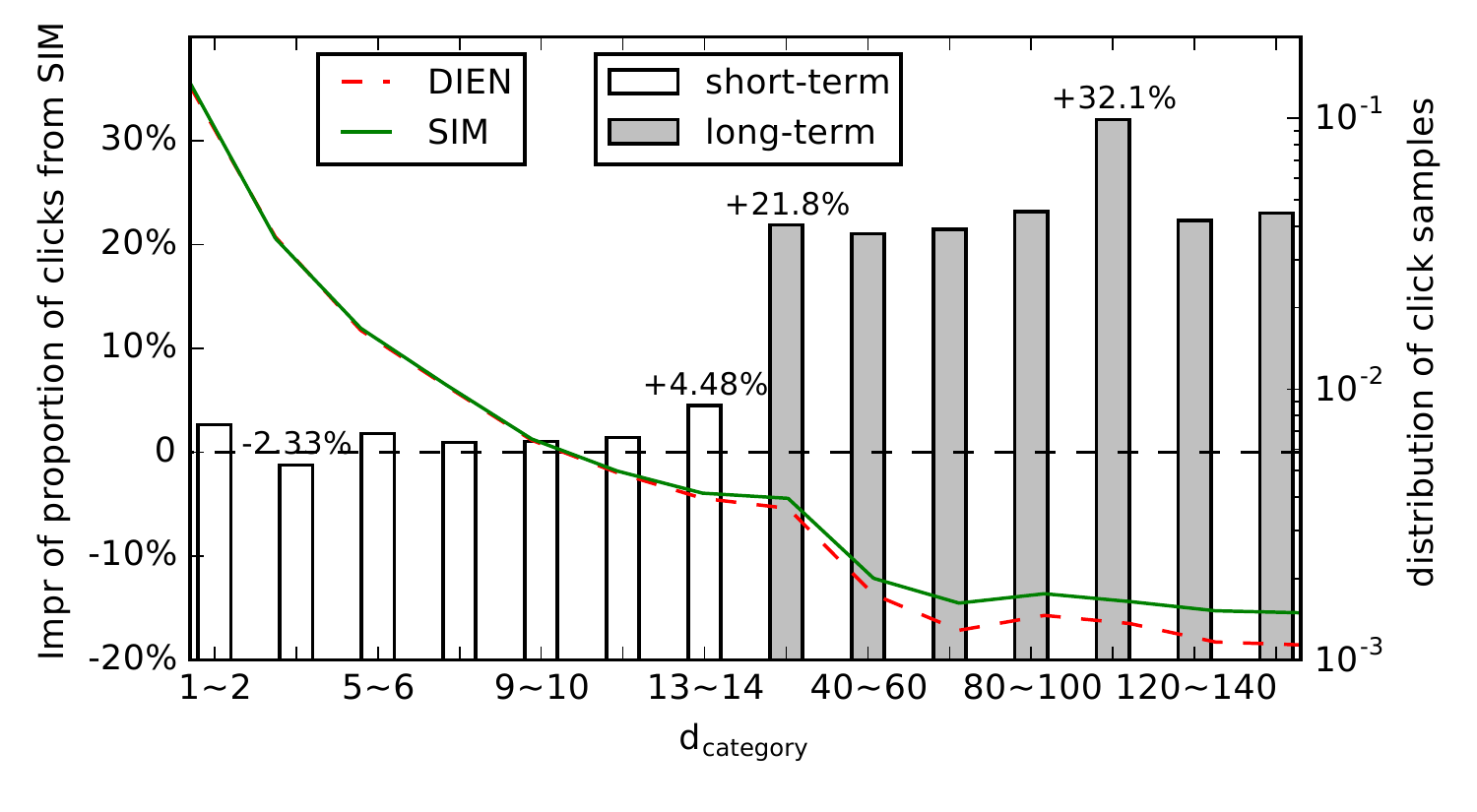} 
        \caption{The distribution of click samples from DIEN and SIM. The clicks are splited into two parts: long-term (>14 days) and short-term ($\leq$14 days), which are aggregated according to the proposed metric Days till Last Same Category Behavior ($d_{category}$). The aggregated scale is different in short-term (2 days) and long-term  (20 days). The boxes demonstrate the lift of proportion of clicks from SIM w.r.t. different $d_{category}$.}
    \label{fig:distribution}
\end{figure}

\textbf{Rethinking Search Model.} We make great efforts on users' long-term interest modeling and the proposed SIM archives good performance on both offline and online evaluation. But does SIM perform better as a result of precise long-term interest modeling? And will SIM prefer to recommend items relevant to people's long-term interest? To answer the two questions, we formulate another metric. \textbf{Days till Last 
Same Category  Behavior}   ($d_{category}$) of a click sample is defined as the days between the users' past behavior on items with the same category as the click sample until the click event happens. For example, user $u_1$ click an item $i_1$ with category $c_1$ and $u_1$ clicked item $i_2$ which has the same category with $i_1$ 5 days ago, and that is $u_1$'s past behavior on $c_1$. If the click event is 
denoted as $s_1$, then the Days till Last Same Category  Behavior of sample $s_1$ will be 5   ($d_{category}^{s_1}=5$). Moreover, if user $u_1$ never has behaviors on category ${c_1}$, we will set $d_{category}^{s_1}$ as $-1$. For a specific model, $d_{category}$ can be used to evaluate the model selection preference on long-term or short-term interest.


After online A/B Testing, we analyze the click samples from SIM and DIEN, which is the last version of the short-term CTR prediction model, based on the proposed metric $d_{category}$. The clicks distribution on $d_{category}$ is shown in Figure~\ref{fig:distribution}. It can be found that there is almost no difference between the two models on short-term part   ($d_{category}<14$) because both SIM and DIEN have short-term user behavior features in the last 14 days. While on the long-term part SIM accounts larger proportion. Moreover, we static the average of $d_{category}$ and the probability of user has historical category behaviors on target item  ($p  (d_{category}>-1)$) on the industrial dataset as shown in Table~\ref{tab:dstatic}. The statistics result on the industrial dataset proves that the improvement of SIM indeed as a result of better longterm interest modeling and compared with DIEN, SIM prefers to recommend items relevant to people's long-term behaviors.
\begin{figure}[t]
    \centering
    \includegraphics[height=1.8in, width=3.5in]{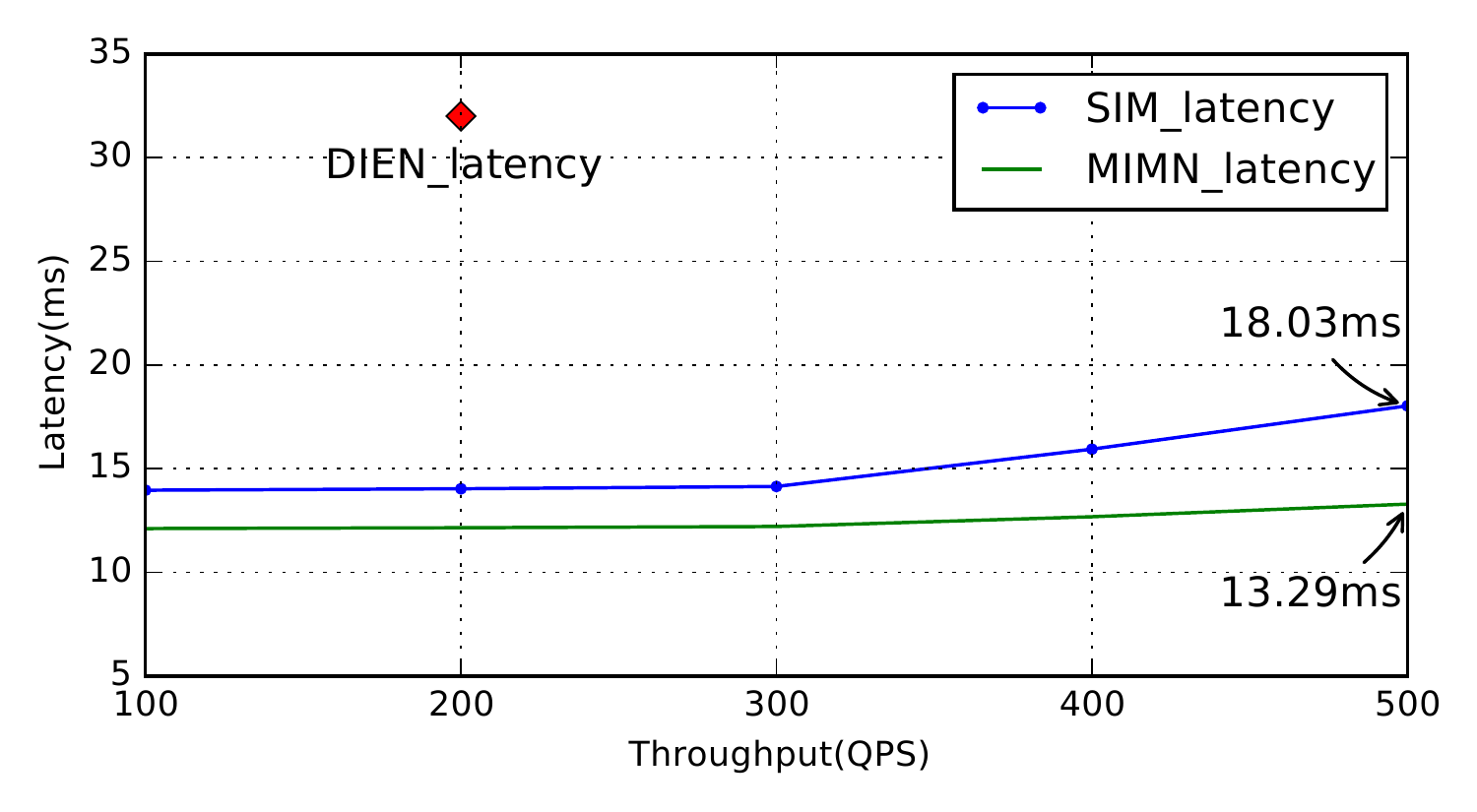} 
    \caption{System performance of realtime CTR prediction system w.r.t. different throughputs. The length of user behavior is truncated to 1000 in MIMN and DIEN, while the length of user behavior can scale up to ten thousand in SIM. The maximum throughputs of DIEN is 200, so there is just one point in the figure.} 
    \label{fig:qps}
\end{figure}

\begin{table}
\caption{SIM's Lift rate of online results compared with MIMN from Jan 7 to Feb 7, 2020, in Guess What You Like column of Taobao App Homepage }\label{tab:online}
            \begin{threeparttable}
                    \begin{tabular}{lcc}
                            \addlinespace
                            \toprule
                         ~~~~~   Metric ~~~~~~ &CTR ~~~~~  ~~~~~  &RPM ~~~~~ \\
                              \midrule
                         ~~~~~   {Lift rate} ~~~~~&$7.1\%$  ~~~~~ & $4.4\%$  ~~~~~ \\
                            \bottomrule
            \end{tabular}
            \end{threeparttable}
\end{table}

\begin{table}
    \caption{Statistics of $d_{category}$ on industrial dataset recommendations}\label{tab:dstatic}
        \begin{tabular}{lcc}
            \addlinespace
            \toprule
        ~~~~    model        ~~       & average $d_{category}$  &$p  (d_{category}>-1)$ \\
            \midrule
        ~~~~    {DIEN}     ~~~~     & $ 11.2 $  &  $ 0.91 $  \\
        ~~~~    {SIM}      ~~~~    & $ \bm{13.3} $  &  $ \bm{0.94} $   \\
            \bottomrule
    \end{tabular}
\end{table}

\textbf{Practical Experience For Deployment.}
Here we introduce our hands-on experiments of implementing SIM in an online serving system. High traffic in Alibaba is well-known, which serves more than 1 million users per second at a traffic peak. Moreover, for each user, the RTP system needs to calculate the predicted CTR of hundreds of candidate items. We build a two-stage index for the whole user behavior data offline, and it will be updated daily. The first stage is the user id. In the second stage, the life long behavior data of one user is indexed by the categories, which this user has interacted with. Although the number of candidate items is hundreds, the number of categories of these items is usually less than 20. Meanwhile, the length of sub behavior sequence from GSU for each category is truncated by 200 (the original length are usually less than 150). In that way, the traffic of each request from users is limited and acceptable. 
Besides, we optimize the calculation of multi-head attention in ESU by deep kernel fusion.

Our real-time CTR prediction system performance of latency w.r.t. throughout with DIEN, MIMN, and SIM show in Figure~\ref{fig:qps}. It is worth noticing that the maximum length of user behavior that MIMN can handle is 1000 and the performance showed is based on the truncated behavior data. While the length of user behavior in SIM is not truncated and can scale up to 54000, pushing the maximum length up to 54x. SIM serving with over ten thousand behavior only increases latency by 5ms compared to MIMN serving with truncated user behavior.

\section{Conclusions}
In this paper, we focus on exploiting over ten thousands of sequential user behavior data in real industrial. Search-based interest model is proposed to capture the diverse user long-term interest with target item. In the first stage, we propose a General Search Unit to reduce the ten thousands of behaviors to hundreds. And in the second stage, an Exact Search Unit utilizes the hundreds relevant behaviors to model the precise user interest. We implement SIM in the display advertising system of Alibaba. SIM brings significant business improvement and is serving the main traffic.

SIM introduces much more user behavior data than the previous methods and the experiment results show that SIM pays more attention on the long-term interest. But the search unit still shares the same formula and parameters among all users. In the future, we will try to build user-specific model to organize life long behavior data of each user with respect to personal conscious. In that way, each user will own their individual model, which keeps modeling the evolving interest of the user.





\bibliographystyle{ACM-Reference-Format}
\balance
\bibliography{SIM}
\end{document}